\documentclass[12pt]{article}
\usepackage{geometry} 
\usepackage{graphicx}
\usepackage[parfill]{parskip}
\usepackage{amssymb}
\usepackage{epstopdf}
\newcommand{\dog}{\,\mathrm{dilog}}
\newcommand{\dd}{\mathrm{d}}
\title{The box integrals in momentum-twistor geometry}
\author{Andrew Hodges\thanks{ andrew.hodges@wadh.ox.ac.uk, http://www.twistordiagrams.org.uk}\\{\footnotesize  {\it Wadham College, University of Oxford, Oxford OX1 3PN, U.K.}}}
\date{19 April 2010} 

\begin{document}

\maketitle

\begin{abstract}
An account is given of how the `box integrals', as used for one-loop calculations in massless field theory, appear in momentum-twistor geometry. Particular attention is paid to the role of compact contour integration in representing the Feynman propagator in twistor space. An explicit calculation of all the box integrals, using only elementary methods, is included.
 
\end{abstract}

\section{Introduction}  

The introduction of {\em momentum-twistor space} in (Hodges 2009) has been greatly developed by Mason and Skinner (2009) and effectively incorporated into the rapid  development by Arkani-Hamed, Cachazo and Cheung (2009) of the powerful Grassmannian formalism for describing scattering amplitudes in supersymmetric gauge theory. This note supplies some detail concerning the way that momentum-twistors can be used to represent the basic box-integrals appearing in the one-loop amplitudes. In particular, a twistor-geometric integral construction of thirty years ago takes on new life when re-interpreted in momentum-twistor space. We shall {\em not} be concerned here with reformulating the actual one-loop amplitudes which are found through the composition of these box-integrals. This, and the extension to higher-order loop integrals, is the subject of very active current work, notably by Arkani-Hamed, Cachazo and their collaborators. This note only discusses  the integration which underlies these investigations. It is closely connected with complementary work by Lionel Mason and David Skinner. Their parallel publication, Mason and Skinner (2010), explores the connection of this integral formalism with the AdS formalism and Wilson loops, and carries it further towards the evaluation of actual loop amplitudes.

\section{The box integrals}

The `box integrals' that concern us are of the form 
$$\int \frac{1}{((p-x_1)^2 + i\epsilon) ((p-x_2)^2 + i\epsilon )((p-x_3)^2 + i\epsilon)((p-x_4)^2 + i\epsilon)} \,\, \dd^4p\, ,$$
where the  $x_1, x_2, x_3, x_4$ are `region space' momentum parameters, characteristic of planar diagrams. These parameters express the conservation of momentum, by $k_{i} = x_{i}-x_{i-1}$, and give rise to the concept of dual conformal symmetry.  It is necessary to treat the non-degenerate case where all the $k_i$ are non-null, and the various degenerate cases when some or all of them are null. 

The $i\epsilon$ in the integral is the conventional indication that the real $p$-integral must be deformed into the complex in such  a way as to avoid the poles of the integrand according to the Feynman prescription. 
But in what follows we shall actually discuss something slightly different and more general, namely   
  \begin{equation}\int \frac{1}{((p-x_1)^2 -\mu^2) ((p-x_2)^2  -\mu^2 )((p-x_3)^2   -\mu^2)((p-x_4)^2  -\mu^2)} \,\, \dd^4p\, ,\label{eqn:boxintegral}
  \end{equation}
where  $\mu$ is a non-zero (complex-valued) parameter with the dimensions of mass. We shall be primarily interested in the limit when $\mu^2\rightarrow 0$ from the correct direction in the complex plane, as this will recover the Feymnan prescription for massless  field theory, but the actual integrals we evaluate will have a non-zero $\mu^2$ parameter. Such integrals are finite. This notation also has the advantage of avoiding confusion with the quite different $\epsilon$ used in dimensional regularization. 

There is a further generalization, in which each factor has a different $\mu_i$ mass parameter. Our basic geometrical setting permits this generalization, but the technical computation of integrals is more complicated. This question is  briefly addressed in Section 8 below. Another reason for a focus on a single parameter parameter $\mu$ is that this should suffice to capture the information which is usually expressed with dimensional regularization methods. Our $\mu^2$-dependent results may be regarded as {\em transforms} of the results as obtained in terms of a  dimensional regularization parameter $\epsilon$.

A further distinction must be drawn between the {\em box integrals} evaluated in this note, and the  {\em box functions} generally tabulated. The box integrals $A$, as functions of momenta, always take the form $A = f/\Delta$, where the numerator $f$ is double-logarithmic and the denominator $\Delta$ is purely algebraic. There are good reasons for regarding  the $f$ as dimensionless multiplying factors, for seeing these factors as capturing the real content of the 1-loop integration, and for tabulating them as the `box functions.' On the other hand, we shall exploit the fact that except in the most degenerate cases, $f$ vanishes when $\Delta$ vanishes, and that  the integral $A$ itself is then finite. So we shall state the results in terms of the integral $A$, thereby including this special situation. The corresponding $f$ can be obtained by a trivial removal of the rational denominator. But there is a subtlety: the box functions as conventionally defined absorb an important extra factor of ${\textstyle \frac{1}{2}}$ which is needed to make the correct connection between the scalar box integrals and the actual loop amplitudes. Thus the box functions $F$ are actually of the form $F={\textstyle \frac{1}{2}} f= {\textstyle \frac{1}{2}} \Delta A.$

\section{Wick rotation and contours in twistor space}

As is very well known, the Feynman contour prescription is equivalent to making a `Wick rotation' of the $p$-integration into the complex, so that the time-component of $p$ runs along the imaginary axis. The integral thus becomes a Euclidean space integral. Adopting this point of view, we start by evaluating (\ref{eqn:boxintegral}) in the special case of coincident $x_i$. Without loss of generality, $x_i = 0$ for each $i$. The box integral becomes
\begin{equation}
\int \frac{1}{(p^2 -\mu^2)^4 } \,\, \dd^4p \, .\label{eqn:boxintegral0}
\end{equation}
Let $p^0 =it, p^1 = x, p^2 =y, p^3 =z$, then take hyperspherical coordinates in $\mathbb{R}^4$ for $(t,x,y,z)$ and the integral can be evaluated as
$$\pm 2\pi^2 \int_{0}^{\infty} \frac{i\, r^3 \dd r}{(r^2 +\mu^2)^4 } \,\, \, = \,\,\, \pm \frac{\pi^2 i}{6(\mu^2)^2},$$
where  the $2\pi^2$ factor comes from the 3-volume of a unit $S^3$. The overall sign depends on contour orientation, equivalent to a choice of Feynman or anti-Feynman prescription, and in what follows we shall neglect it.

The twistor translation of this idea turns out to give a striking example of twistor geometry. However, the first step only goes half-way to twistor space; we only go as far as the representation of complexified Minkowski space $\mathbb{CM}$ by  $\mathbb{CP}^5$. Of course, this representation involves the {\em compactification} of  $\mathbb{CM}$, and its transformation under conformal transformations, which we shall address shortly. 

The underlying relations between the conformal group $\mathrm{C}(1,3)$, the $\mathrm{SO}(2, 4)$ acting on $\mathbb{CP}^5$, and the $\mathrm{SL}(4, \mathbb{C})$ acting on twistor space, go back to the very origins of twistor theory in Roger Penrose's first  work.  The reader is referred also to the paper by Mason and Skinner (2009), which gives an introduction to twistor geometry using $\mathbb{CP}^5$ as a bridge from Minkowski space. For present purposes, the most important feature of the correspondence is that  the points in $\mathbb{CM}$ are represented by {\em simple} skew bi-twistors $Q^{\alpha\beta}$  in $\mathbb{CP}^5$, i.e.\ $Q^{\alpha\beta}$ such that $Q^{\alpha\beta}= A^{[\alpha}B^{\beta]}$ for some twistors $A^{\alpha}, B^{\beta}$. This condition  is equivalent to $\epsilon_{\alpha\beta\gamma\delta}Q^{\alpha\beta}Q^{\gamma\delta} = 0$, or to the {\em rank} of $Q$, considered as a linear transformation, being 2.  Geometrically, this condition characterises the {\em Klein quadric} on $\mathbb{CP}^5$. Points of $\mathbb{CP}^5$ not on the quadric correspond to complexified spheres on $\mathbb{CM}$. These are essentially the very  hypersurfaces of form $(p-x)^2- \mu^2=0$ which appear as poles in our integral. As we shall see, the linear structure on $\mathbb{CP}^5$ is highly advantageous in the handling of these poles.

As $\epsilon_{\alpha\beta\gamma\delta}$ acts as an inner product structure, it is natural to introduce the notation $X.Y$ for ${\textstyle \frac{1}{2}} \epsilon_{\alpha\beta\gamma\delta} X^{\alpha\beta}Y^{\gamma\delta}.$ We shall also need the special element $I^{\alpha\beta}$, corresponding to the vertex of the null cone at infinity, and the origin bi-twistor  $O^{\alpha\beta}$. The infinity bi-twistor  $I^{\alpha\beta}$ is defined non-projectively, and knowing its scale corresponds to knowing the metric on $\mathbb{CM}$. The relation is given by:
\begin{equation} (x_1-x_2)^2 = -2\, \frac{X_1.X_2}{(I.X_1)(I.X_2)} \, .
\end{equation}
We may let simple skew bi-twistors $X_1^{\alpha \beta}, X_2^{\alpha \beta}, X_3^{\alpha \beta}, X_4^{\alpha \beta}$ correspond to the points $x_1, x_2, x_3, x_4$ in region space.
We may also regard the variable $p$ as an internal region and give it bi-twistor coordinates $P^{\alpha \beta}$. In the case $x_i = 0$, as studied above, the integrand of (\ref{eqn:boxintegral0})   then translates into
$$ \frac{(I.P\, O.I)^4}{((2 O + O.I \mu^2 I).P)^{4}}$$
The bi-twistor $(2O^{\alpha\beta}+O.I \mu^2 I^{\alpha\beta})$ is an example of a {\em non}-singular element of the $\mathbb{CP}^5$.
  
 So far it would appear that twistors are being used only to express these bi-twistors; equivalently, only  the {\em lines} of projective twistor space actually play a role, and not the points or planes. Moreover, since the box integral is a completely scalar structure, there might appear to be no reason for the spin-structure implicit in twistor space to be relevant. It is only when we seek to translate the differential form and the contour in (\ref{eqn:boxintegral0}), that we find twistor space coming naturally into the picture.

Suppose we do indeed resolve a bi-twistor into two representative twistors, and so write $P^{\alpha\beta} = Z^{\alpha}V^{\beta}-V^{\alpha}Z^{\beta} $ where $ Z^{\alpha}= (ip^{AB'}\sigma_{B'}, \sigma_{A'}),$ $ V^{\beta} = (ip^{BA'}\tau_{A'}, \tau_{B'})$. 

Then  $\dd^4Z \wedge \dd^4V =  i (\sigma_{A'}\tau^{A'})^{2} \dd^4p \wedge \dd^2\sigma \wedge \dd^2 \tau  $, where $\sigma_{A'}\tau^{A'}$ is the same thing as $I_{\alpha\beta}Z^{\alpha}V^{\beta}.$ (Note that here and in what follows we shall use {\em non-projective} twistor spaces and differential forms  for the calculations, but this is only conventional; at each stage the integrals could be re-expressed in projective spaces if desired.)

Now consider the eight-dimensional integral
\begin{equation}\int \frac{\dd^4Z\wedge \dd^4V}{(2 (O.I)^{-1}O_{\alpha\beta}+ \mu^2 I_{\alpha\beta})Z^{\alpha}V^{\beta})^{4}}  
= \int \frac{i\, \dd^4p \wedge \dd^2\sigma \wedge \dd^2 \tau}{(p^2 -\mu^2)^4(\sigma_{A'}\tau^{A'})^2}
\, . \label{eqn:feynmanbase}
\end{equation}
where the contour is defined by letting  $p$ run  over the (non-compact) $\mathbb{R}^4$ as defined above, and the spinor integral to be taken independently of $p$ over  an $S^3 \times S^1$. This may be done by taking an `antipodal' or `anti-diagonal' contour where  $\tau_{A'} = t^{AA'}\bar{\sigma}_A$, for some Hermitian non-singular $t^{AA'}.$  It gives a factor $(2\pi i)^3.$ Thus the result (up to sign) is
$$\frac {\pi^2  }{6( \mu^2)^2}(2\pi i)^3 \, ,$$
yielding a representation of the Feynman integral by a twistor integral. But
it might be considered that genuine {\em twistor} structure has still not played any part. What matters in the integral is only the integration over the  $P^{\alpha\beta}$. The spinor integral is a trivial factor; it merely integrates over the different ways in which  $P^{\alpha\beta}$ is represented as a skew product of a $Z^{\alpha}$ and a $V^{\beta}$, and these representations do not have any significance. We seem to have added four extra dimensions, and then to have integrated them out, for no purpose.

Twistor-geometric structure emerges only when we {\em compactify} this integral.  
To achieve this compactification, first let $t^{AA'}$ be the unit vector in the 0-direction. 
We can then define the {\em Euclidean dual} of a twistor $Z^{\alpha} = (\omega^A, \pi_{A'})$ to be the dual twistor $\tilde{Z}_{\alpha} = (t^{AA'}\bar{\pi}_{A}, t_{AA'}\bar{\omega}^{A'}).$

We can also define the {\em Euclidean norm} of a twistor $Z^{\alpha} = (\omega^A, \pi_{A'})$ to be given by $|Z| = \sqrt{Z^{\alpha} \tilde{Z}_{\alpha}} = \sqrt{\omega^A\bar{\omega}^{A'}t_{AA'} + \pi_{A'}\bar{\pi}_{A}t^{AA'}}. = \sqrt{|\omega^0|^2 +|\omega^1|^2+ |\pi_{0'}|^2 + |\pi_{1'}|^2}\, .$

Next we consider the space of normalised twistors described by: $$ Z^{\alpha}= \frac{(ip^{AB'}\sigma_{B'}, \sigma_{A'})}{|(ip^{AB'}\sigma_{B'}, \sigma_{A'})|}\, ,$$
where $p$ runs over the same space as before.
These satisfy $|Z|^2 = |\omega^0|^2 +|\omega^1|^2+ |\pi_{0'}|^2 + |\pi_{1'}|^2 = 1$, but do not fill out a complete $S^7$, because the twistors of form $(\rho^{A}, 0)$ are absent. These, of course, belong to the null cone at infinity in $\mathbb{CM}$. Now note that provided $p \ne 0, \hat{p} \ne 0,$ we also have
 $$ Z^{\alpha}= \frac{(\rho^{A}, i\hat{p}_{AA'}\rho^{A})}{|(\rho^{A}, i\hat{p}_{AA'}\rho^{A})|}$$
 where $\hat{p}^{AA'} = p^{AA'}/p^2$, and $\rho^{A} = ip^{AB'}\sigma_{B'}$. So we may identify $ p \ne 0, \hat{p} \ne  0$ as defining two coordinate patches, together covering an extended contour. The point $\hat{p}=0$ gives a one-point compactification of the original Euclidean $\mathbb{R}^4$, turning it  into an $S^4$. Moreover, the addition of the twistors of form $(\rho^{A}, 0)$ means that the seven-dimensional space of normalised $Z$ now fills out the complete $S^7$. For every projective twistor in $\mathbb{CP}^3$, there is an $S^1$ of points in this $S^7$, thus realizing the Hopf fibration. 
 
 Note that the Euclidean dual of $Z^{\alpha}$ is $$ \tilde{Z}_{\alpha}= \frac{( t^{AA'}\bar{\sigma}_{A}, -it_{AA'}\bar{p}^{A'B}\bar{\sigma}_{B})}{|(ip^{AB'}\sigma_{B'}, \sigma_{A'})|}\, ,$$
 but that for $p$ of our form, $\bar{p}$ coincides with the time-reversed $p$, which means that $\bar{p}^{A'B} = p^{A'B} - 2(p.t)t^{A'B} = t^{AA'}p_{AB'}t^{BB'}$. Hence 
  $$ \tilde{Z}_{\alpha}= \frac{( t^{AA'}\bar{\sigma}_{A}, -i{p}_{A}^{B'}t^B_{B'} \bar{\sigma}_{B})}{|(ip^{AB'}\sigma_{B'}, \sigma_{A'})|}\, .$$
 If we further define $$V^{\alpha} = (O^{\alpha\beta}  +  I^{\alpha\beta}) \tilde{Z}_{\beta}$$
 then we have
 $$V^{\alpha} = \frac{(ip^{AB'}t^B_{B'} \bar{\sigma}_{B},  t^{AA'} \bar{\sigma}_{A})}{|(ip^{AB'}\sigma_{B'}, \sigma_{A'})|}\, .$$
 
The pair $(Z, e^{i\theta}V)$ then gives an eight-dimensional set with $S^7 \times S^1$ topology.  Integrating over the coordinate patch $\hat{p}\ne 0$ amounts to exactly the same as  (\ref{eqn:boxintegral0}). The remaining points, with $\hat{p} = 0$, are only a set of measure zero and  make no difference to the value of the integral.  Thus, integration over this complete compactified contour is equivalent to the Feynman integration. But the topology is completely changed by the compactification: in the compact contour, the spinor integral does {\em not} factorise trivially, and the integration is truly twistor-geometric.

Having defined this compactification of the contour, it is much simpler to work in twistor space without any reference to integration in Minkowski space. Indeed this was the original construction in (Hodges 1977), as described in the historical note in section 10 below.
The starting-point then was the fundamental twistor integral, as described for instance in (Penrose and McCallum 1972):
\begin{equation}\frac{1}{(2\pi i)^5}\, \oint \frac{6\,  \dd^4Z\wedge \dd^4W}{(Z^{\alpha}W_{\alpha})^{4}} = 1\, ,\label{eqn:fundamental} \end{equation}
where the contour can be represented by a $S^7 \times S^1$. By the coordinate change $W_{\alpha}=Q_{\alpha\beta}V^{\beta},$ we have
$$\frac{1}{(2\pi i)^5} \oint \frac{6\, \dd^4Z\wedge \dd^4V}{(Q_{\alpha\beta}Z^{\alpha}V^{\beta})^{4}}  = \frac {1}{\det Q} \, ,$$
where $Q$ is any non-singular linear transformation. If $Q$ is antisymmetric, the determinant simplifies to the square of the Pfaffian and we have
\begin{equation}\frac{1}{(2\pi i)^5} \oint \frac{6\, \dd^4Z\wedge \dd^4V}{(Q_{\alpha\beta}Z^{\alpha}V^{\beta})^{4}}  = \frac {16}{(\epsilon^{\alpha\beta\gamma\delta}Q_{\alpha\beta}Q_{\gamma\delta})^2}\, .\label{eqn:pfaff} \end{equation}
In particular, for $Q^{\alpha\beta}= 2 (O.I)^{-1}O^{\alpha\beta}+ \mu^2 I^{\alpha\beta}$, we obtain
\begin{equation} \frac{1}{(2\pi i)^5} \oint \frac{6\,  \dd^4Z\wedge \dd^4V }{((2 (O.I)^{-1}O_{\alpha\beta} + \mu^2 I_{\alpha\beta})Z^{\alpha}V^{\beta})^{4}}= \frac {16}{(8 \mu^2)^2}\label{eqn:basic}\end{equation}
in agreement with (\ref{eqn:feynmanbase}). This is the fundamental compact twistor integral on which everything that follows rests.

An important feature of the compactification is that the non-singularity of the form on the  compactified contour is equivalent to the  {\em absence of an ultra-violet divergence}.

When the Feynman integration over momentum $p$ is recast as this compact twistor integration of a holomorphic form, it becomes free from its original definition in terms of  spacetime. The Feynman propagator is thus given a new geometrical interpretation, which is essentially defined by the $\mathbb{CP}^3$ of projective  twistor space. It is a striking fact that from this point of view,  a loop integral has the shape of twistor space.
A further freedom arises from the fact that the twistor-space contour,  more correctly regarded as a homology class, need not possess any immediately obvious connection with the topology of an $S^7$.  In  section 10, an example is given of this freedom of representation.

The underlying structure is the quaternionic fibration of $S^7$ over $S^4$, analogous to the Hopf fibration. That is, we consider an $S^7$ as the space of pairs of quaternions $(q_1, q_2)$ with $|q_1|^2 + |q_2|^2 =1$. Then define equivalence classes on $S^7$ by $(q_1, q_2) \sim (q'_1, q'_2)$ if $q'_1 = qq_1, q'_2= qq_2$ for some unit-normed $q$. The quotient space is homeomorphic to $S^4$ and each fibre is an $S^3$. This fibration rests on the fact that the quaternionic norm satisfies $|qr| = |q||r|$, which itself can be expressed simply as the four-square identity for real numbers.\footnote{ 
I am indebted  to Lionel Mason for pointing out that the $S^7\times S^1$ contour must be connected with the Euclidean space by this fibration. Earlier work had used the $S^7\times S^1$, noting that it gave the correct answer, without seeing the underlying reason for this correctness.}

\section{Generalised Feynman parameters}

From now on we can abandon the original $p$-space integration, and derive everything from the compact twistor contour integral:
\begin{equation}\frac{1}{(2\pi i)^5} \oint \frac{6\, \dd^4Z\wedge \dd^4V }{(Q_{\alpha\beta}Z^{\alpha}V^{\beta})^{4}} = \frac {16}{(\epsilon^{\alpha\beta\gamma\delta}Q_{\alpha\beta}Q_{\gamma\delta})^2}\, . \label{eqn:twistorintegral}
\end{equation}
This is  valid for all non-singular anti-symmetric  $Q$ in $\mathbb{CP}^5$, i.e.\ those which are not on the quadric  $\epsilon^{\alpha\beta\gamma\delta}Q_{\alpha\beta}Q_{\gamma\delta}=0$ which corresponds to complexified Minkowski space.  The difficulty in our programme, of course, is that we are interested precisely in the limit $\mu^2 \rightarrow 0$ where the limiting $Q$ {\em is} on this quadric, so that this compact twistor contour no longer exists. 

We now address the original integral, with its four different $x_i$.  Rather than find a new contour in twistor space, we think of moving the $x_i$ apart from coincidence, while keeping the twistor contour the same.  There are very well known techniques for doing just this, using {\em Dirichlet averaging,} in the form used extensively by  Feynman.
The principle is that of embedding the integral in a larger space and then exchanging the order of integration. Feynman's classic method uses the identity expressed in a symmetrical form by
$$\frac{1}{Q_1Q_2Q_3Q_4} = \int_0^{1}\!\int_0^{1}\!\int_0^{1}\!\int_0^{1}\!\frac{6 \, \delta(1-\alpha_1 -\alpha_2 -\alpha_3 - \alpha_4)}{(\alpha_1 Q_1 + \alpha_2 Q_2 + \alpha_3 Q_3 + \alpha_4 Q_4)^4} \dd\alpha_1\dd\alpha_2\dd\alpha_3\dd\alpha_4 $$
which is an
 integration over a tetrahedron in $\mathbb{R}^3$. But we are at liberty to prefer 
the identity:
\begin{equation}\frac{1}{Q_1Q_2Q_3Q_4} = \int_0^{\infty}\int_0^{\infty}\int_0^{\infty} \frac{6\, \dd\alpha \,\dd\beta \,\dd\gamma}{(Q_1+\alpha Q_2+\beta Q_3 + \gamma Q_4)^4} \, .\label{eqn:cube} \end{equation}
Here and throughout, integration to $\infty$ is to be interpreted as a compact integral in a suitable (complex) projective space; so in this formula, $\alpha, \beta, \gamma$ are actually $\mathbb{CP}^1$ parameters. The topology of the space is that of a {\em cube} in $(\mathbb{CP}^1)^3$. This may be considered to correspond with the Feynman tetrahedron by $$\alpha= \alpha_1(1-\alpha_1-\alpha_2-\alpha_3)^{-1}, \beta= \alpha_2(1-\alpha_1-\alpha_2-\alpha_3)^{-1}, \gamma= \alpha_3(1-\alpha_1-\alpha_2-\alpha_3)^{-1}\, ,$$ 
a mapping which blows up the $\alpha_4=1$ face into the three faces of a cube given by $\alpha=\infty, \beta=\infty, \gamma=\infty$. Clearly, this parameter space breaks the symmetry which is respected by the Feynman tetrahedron, but as a direct product of three line intervals it may be more convenient for computational purposes.
We shall also use a 3-volume with the shape of a {\em triangular prism}, and the identity 
\begin{equation}\frac{1}{Q_1Q_2Q_3Q_4} = \!\!\int_0^{\infty}\!\!\!\dd x\! \int_{\Delta}\frac{6 (1+x)^2(1 - \alpha_1 - \alpha_2) \quad \dd\alpha_1\dd\alpha_2}{( (1+x)(\alpha_1 Q_2 + \alpha_2 Q_3) + (1 - \alpha_1 - \alpha_2)(xQ_1 + Q_4))^4} \, .\label{eqn:prism} \end{equation}
where $\Delta$ is the triangular region $0 \le \alpha_1, 0 \le  \alpha_2, \alpha_1 + \alpha_2 \le 1$. This can be regarded as arising from taking the Feynman tetrahedron and replacing  $\alpha_3$ by $x=\alpha_3 (1-\alpha_1-\alpha_2-\alpha_3)^{-1}$ as a parameter. This blows up one of the edges of the tetrahedron into a rectangular face. 

Such transformations of the Feynman parameters are nothing new, and indeed no new results will follow from the analysis in this note. The most general integral of this form, with different $\mu_i$, was calculated exactly by t'Hooft and Veltman (1979) long ago. We are merely reviewing the derivation of the main results in the new context provided by the compact twistor-space contour and the  use of  the linear structure of $\mathbb{CP}^5$ instead of that of $\mathbb{R}^4$.

The Feynman parameter method removes the need to do any integration over the original space (here replaced by a twistor space), once the fundamental integral has been done.
If both the Feynman parameter space and the twistor contour are {\em compact}, then the change of order of integration can be rigorously justified. However, this depends upon the {\em same} twistor contour being used for each parameter point, and this is harder to establish. The fact that the twistor contour is  compact gives a very useful starting-point. Having established a specific contour for (\ref{eqn:twistorintegral}), for some specific $Q_0$, the same contour must be valid for $Q$ in some open neighbourhood of $Q_0$, and in particular for a polyhedron in that neighbourhood. Restricted to  this region, therefore, the method is rigorous. In practice we extend to {\em all} non-singular $Q$ by an argument from analytic continuation, thus assuming in effect that the contour can always be moved around as required in the twistor space. Indeed we can go further and by moving $Q$ parameters in a loop, deduce the existence of period contours. Whilst more  careful mathematical study would be desirable, we are not seriously concerned about the validity of the results obtained by informal methods.

\section{Evaluation of the 4-mass case}

The `4-mass' box integral is simply the non-degenerate case where all the external $K_i$ are non-null, and so equivalently, all the $x_i$ are non-null separated. In $\mathbb{CP}^5$ language, the corresponding elements satisfy  $X_i.X_j \ne 0$. The corresponding lines in twistor space are all {\em skew.} We now wish to evaluate the box integral (\ref{eqn:boxintegral}) in this case, and in the limit as $\mu^2\rightarrow 0$.

We have established that by integrating over the direct product of a compact $S^7 \times S^1$ contour in the twistor space, and a cube-shaped region in the Feynman parameter space,
\begin{eqnarray}& &\frac{1}{(2\pi i)^5} \oint \frac{6\,  \dd^4Z\wedge \dd^4V }{(Q_{1\alpha\beta}Z^{\alpha}V^{\beta})(Q_{2\alpha\beta}Z^{\alpha}V^{\beta})(Q_{3\alpha\beta}Z^{\alpha}V^{\beta})(Q_{4\alpha\beta}Z^{\alpha}V^{\beta})}
\nonumber \\
&&=
\int_0^{\infty}\!\!\int_0^{\infty}\!\! \int_0^{\infty} \!\! \frac {\dd\alpha\, \dd\beta \,\dd\gamma}{\det (Q_1+\alpha Q_2 + \beta Q_3 + \gamma Q_4)} . \label{eqn:qintegral} \end{eqnarray}
So we may apply this to the case where
$Q_i^{\alpha\beta}= 2(X_i.I)^{-1}X_i^{\alpha\beta} + \mu^2I^{\alpha\beta}$, and the $X_i$ correspond to four skew lines in twistor space.
Using the abbreviation $x_{ij}^2$ for  $-2(X_i.X_j)(X_i.I\, X_j.I)^{-1}$, the result  is
$$\int_0^{\infty}\!\!\int_0^{\infty}\!\! \int_0^{\infty} \!\! \frac{ \dd\alpha\, \dd\beta\, \dd\gamma}{(\alpha x_{12}^2 + \beta  x_{13}^2 + \gamma  x_{14}^2 + \alpha \beta  x_{23}^2 + \alpha \gamma  x_{24}^2 + \beta\gamma  x_{34}^2 -\mu^2(1+\alpha+\beta+\gamma)^2)^2} \, ,$$
where the connection with the standard kinematic parameters is given by  $$x_{13}^2=t,\, x_{24}^2=s, \, x_{14}^2 = K_1^2, \, x_{12}^2 = K_2^2, \, x_{23}^2 = K_3^2, \, x_{34}^2 = K_4^2\, .$$
Here we take a short cut by taking the limit $\mu^2 \rightarrow 0$ inside the integral. In doing so, we  are  neglecting to take proper account of how the limit $\mu^2 \rightarrow 0$ encodes the Feynman prescription. In effect, we shall find the amplitude function as a many-branched complex function, leaving to later the question of which branch is the correct one to take. 

The $\gamma$ integration is trivial and leaves
$$\int_0^{\infty}\int_0^{\infty}   \frac {\dd\alpha \,\dd\beta}{(\alpha x_{12}^2 + \beta x_{13}^2 + \alpha\beta x_{23}^2)( x_{14}^2 +  \alpha x_{24}^2 + \beta x_{34}^2)}  \, .$$
The $\alpha$ and $\beta$ parameters admit a simple rescaling
$u = \alpha x_{24}^2/x_{14}^2, \, v = \beta x_{34}^2/x_{14}^2,$
which transforms the integral into
$$\int_0^{\infty}\int_0^{\infty}  \frac { \dd u\, \dd v}{(ux_{12}^2x_{34}^2  + vx_{13}^2x_{24}^2  + uv x_{23}^2x_{14}^2 )( 1 + u +v )} $$
so that evaluation amounts to performing the double integral
\begin{equation}\int_0^{\infty}\int_0^{\infty}   \frac {\dd u\, \dd v}{(auv + bu  + cv  )( 1 + u +v )} \, ,\label{eqn:uv}\end{equation}
where $a=x_{23}^2x_{14}^2= K_3^2K_1^2, \, b =x_{12}^2x_{34}^2 = K_2^2K_4^2, \, c = x_{13}^2x_{24}^2= st$. 
 Note that the ratios $a/b, b/c, c/a$ are conformal invariants. 
 
 Performing the integration over $v$ leaves
\begin{equation}\int_0^{\infty} \frac{\log((au+c)(1+u)/bu)}{au^2 + (a+c-b)u + c} \, \dd u\, .\label{eqn:uacintegral} \end{equation}
We note that
$$\int_0^{\infty} \frac{\log((au+c)/bu)}{au^2 + (a+c-b)u + c} \dd u = \int_0^{\infty} \frac{\log((w+1)a/b)}{aw^2 + (a+c-b)w + c} \dd w$$
 by $w = c/au$, so that the integral (\ref{eqn:uacintegral}) is equal to
\begin{equation}\int_0^{\infty} \frac{2\log(1+u) + \log(a/b)}{au^2 + (a+c-b)u + c}\,  \dd u \, .\label{eqn:uintegral}\end{equation}
In what follows we shall write $(-\kappa)$ and $(-\tilde{\kappa})$ for the two roots of the quadratic $au^2 + (a+c-b)u + c$, so that $\kappa\tilde{\kappa} = c/a$ and $\kappa + \tilde{\kappa} = (a+c-b)/a $. We also write $\Delta$ for $ a(\kappa -   \tilde{\kappa})$, so that $\Delta^2 = (a+c-b)^2 - 4ac = a^2+b^2+c^2-2ab-2bc-2ca$.

In these terms the integral (\ref{eqn:uintegral}) becomes
$$ \int_0^{\infty}
\frac{(2\log(1+u) + \log(a/b))}{a(u+\kappa)(u+ \tilde{\kappa})} \dd u\, = \frac{1}{2\pi i}  \oint 
\frac{(2\dog(1+u) + \log(a/b)\log u)}{a(u+\kappa)(u+ \tilde{\kappa})} \dd u\, ,$$ 
by using the property of a contour surrounding the dilogarithmic cut (see equation (\ref{eqn:cut}) in the Appendix, where some basic properties of the dilogarithm are listed).
Deforming the contour into circles around the two poles, we finally evaluate the integral as:
\begin{eqnarray} 
&&\frac{2\dog(1-\kappa) - 2\dog(1 - \tilde{\kappa}) + \log(a/b)\log ( \kappa/\tilde{\kappa})}{\Delta}\nonumber \\
&=&\frac{2\dog(1-\kappa) - 2\dog(1 - \tilde{\kappa}) - \log((1-\kappa)(1-\tilde{\kappa}))\log ( \kappa/\tilde{\kappa})}{\Delta}\, .\label{eqn:dilogform}\end{eqnarray} 
This can be written in many other forms, because any permutation of $(x_1, x_2, x_3, x_4)$ and hence of $(a, b, c)$ corresponds to a cross-ratio transformation generated by $\kappa \rightarrow 1 - \kappa, \kappa \rightarrow  \kappa^{-1}.$  The dilogarithm is  invariant, up to logarithmic terms, under such cross-ratio transformations, and so numerous identities are available. Using one such dilogarithmic identity (\ref{eqn:zz-1}) it can immediately be written as
  $$-\frac{2\dog(1-\kappa^{-1}) - 2\dog(1 - \tilde{\kappa}^{-1}) - \log((1-\kappa^{-1})(1-\tilde{\kappa}^{-1}))\log ( \kappa/\tilde{\kappa})}{\Delta} $$ which is equivalent to the exchange of $a$ and $c$.
Averaging over these two expressions gives a longer formula which is equivalent to the expression preferred by  Bern et al.\ (2004), in their equation (41).

Another expression, longer but more elegant and manifestly invariant under permutations of $(x_1, x_2, x_3, x_4),$ is given by \begin{equation}\frac{1}{\Delta}\left\{ \begin{array}{l}\dog(\kappa/\tilde{\kappa}) - \dog(\tilde{\kappa}/ \kappa) - \dog(1-\kappa/1- \tilde{\kappa}) + \dog(1-\tilde{\kappa}/ 1- \kappa)\\+ \dog(1-\kappa^{-1}/1- \tilde{\kappa}^{-1}) - \dog(1-\tilde{\kappa}^{-1}/ 1- \kappa^{-1})\end{array}\right\}\, .\label{eqn:kappa}\end{equation}
To establish this, write (\ref{eqn:uintegral}) as
$$ \int_0^{\infty}
\frac{\log((1+u)/(1-\kappa) )+ \log((1+u)/(1-\tilde{\kappa})}{a(u+\kappa)(u+ \tilde{\kappa})} \, \dd u\, .$$ 
For the first logarithm, change the contour to run from 0 to $-\kappa$, where there is a removable singularity, and then from  $-\kappa$ to $\infty$, changing variables by a M\"{o}bius transformation to $z=-(u+\kappa)/(u+\tilde{\kappa})$. For the second logarithm,  exchange the roles of $\kappa$ and $\tilde{\kappa}$. The result is then immediate. This use of a M\"{o}bius transformation based on a removable singularity is  equivalent to proving Abel's fundamental identity for the dilogarithm. 

A special case arises if the roots are coincident, so $\Delta=0$. This condition, equivalent to $\sqrt{a} \pm \sqrt{b} \pm \sqrt{c} = 0$, is closely analogous to Ptolemy's condition (that $|AB||CD|  \pm |AC||BD| \pm |AD||BC| = 0$) for four points $A, B, C, D$ in the Euclidean plane to be cocylic. Geometrically, $\Delta=0$ implies that the four points lie on a complexified circle, or that that the four elements of $\mathbb{CP}^4$ are linearly dependent. In this case the integral is still finite, and its value can be written symmetrically as
$$\frac{\log a}{a-b-c} +\frac{\log b}{b-c-a} + \frac{\log c}{c-a-b} \, .$$

From the $uv$-integral (\ref{eqn:uv}) one may read off the existence of three {\em period} contours, obtained by analytically continuing in $a, b, c$. These are given by 
\begin{eqnarray}&&2\pi i\int_0^{\infty} \frac{1}{au^2 + (a+c-b)u + c} \dd u \, , \quad 2\pi i\int_0^{\infty} \frac{1}{bu^2 + (b+a-c)u + a} \dd u \, ,\nonumber \\ && 2\pi i\int_0^{\infty} \frac{1}{cu^2 + (c+b-a)u + b} \dd u\, ,
\label{eqn:logperiod} \end{eqnarray}
which yield 
$$2\pi i\frac{\log(\kappa/\tilde{\kappa})}{\Delta}, \quad 2\pi i\frac{\log(1-\kappa/1-\tilde{\kappa})}{\Delta}, \quad 2\pi i\frac{\log(1-\kappa^{-1}/{1-\tilde{\kappa}^{-1})}}{\Delta} \, ,$$
again with special cases when $\Delta = 0$.
These three contours add to zero, modulo the {\em double} period which gives $(2\pi i)^2 \Delta^{-1}$.
It is the double period which corresponds to the `leading singularity' of the amplitude, and also corresponds to the integral obtained by changing the Feynman propagator to a $\delta$-function on the light-cone, i.e.\ putting the propagator `on-shell'. 
This `leading singularity' structure is very significant in the current development of loop amplitude theory. Analogous phenomena of period contours have appeared in twistor-theoretic literature since the very earliest days, though in a different physical context. A historical note is given in section 10. 

This period structure is relevant to a question which we have not yet addressed, namely that of how the Feynman prescription is translated into the correct choice of twistor contour, and thus into the correct branch of the dilogarithm. Here the work of Duplan\u{c}i\'{c} and Ni\u{z}i\'{c} (2002) emphasises a subtle breaking of conformal invariance in the amplitude. Although the dilogarithmic function, as a complete analytic function, is a function only of the conformally invariant ratios of $a/b, b/c$, the choice of which
branch to take depends on knowing the individual variables $s, t, K_i^2$. Thus, the $\mu^2$ parameter leaves behind a remnant of the infra-red divergence it regularises, even when the limit is finite. 
This is evident from the original definition of the integral, in which the Feynman prescription applies to the individual $x_i$. In the derivation of the dilogarithmic function we have neglected this prescription in two places: first when the limit $\mu^2=0$ is taken before doing the integral, and then in the rescalings  $u = \alpha x_{24}^2/x_{14}^2, v = \beta x_{34}^2/x_{14}^2$. One may start with a configuration where all the $x_i$ are spacelike-separated; in this case there is  no ambiguity in the integral, nor in the rescalings, and  $u, v, w$ may all  be taken to run along the real positive axis. But analytic continuation of the resulting amplitude to timelike-separated $x_i$ requires knowledge of the Feynman prescription. Duplan\u{c}i\'{c} and Ni\u{z}i\'{c} express the choice of branch in terms of rules for adding on a logarithmic period function.

It is worth noting that Duplan\u{c}i\'{c} and Ni\u{z}i\'{c} (2002) make use of the {\em triangle} integral in order to make their definition of the correct dilogarithmic branch.  In our picture this triangle integral also has a natural geometric meaning: it simply corresponds to putting $Q_4 =I$ in (\ref{eqn:qintegral}).

\section{Momentum-twistor parameters and \newline transversals}

So far we have  expressed the external parameters $x_i$ in terms of elements $X_i$ of $\mathbb{CP}^5$. But we can also separate these $\mathbb{CP}^5$ elements into representative twistors. They can immediately be identified as the momentum-twistors as introduced in (Hodges 2009).

One motivation for doing this is that when the box-integrals are combined with the helicity structure of actual gauge-fields, we shall certainly need such helicity-carrying twistor parameters. But even in the 4-mass scalar box-integral, where no helicity structure is apparent, the twistor-space structure is a useful adjunct. 

In (projective) twistor space, the four external $x_i$ will correspond to four skew lines.
The analysis of the 4-mass integral has shown that it behaves as a residue calculation, the poles being determined by $\kappa$ and $\tilde{\kappa}$ as the solutions of a (conformally invariant) quadratic equation. This structure reflects the existence of just two {\em transversals} to those four skew lines. The special case 
$\kappa = \tilde{\kappa}$ corresponds to the coincidence of those two transversals.

\begin{figure}[h] 
   \centering
   \includegraphics[width=181px]{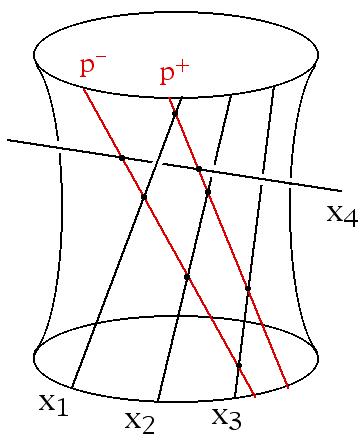} 
    \caption{4-mass transversals}
   \label{fig:Fig.1}
\end{figure}

In Minkowski space, the transversals correspond to the solutions $p^{\pm}$ of the equations
$(p-x_1)^2=(p-x_2)^2= (p-x_3)^2= (p-x_4)^2 = 0.$
In principle this is just a quadratic equation, but it is surprisingly difficult to write down a formula for $p^{\pm}$. The reason is that giving a formula in Minkowski coordinates involves a fifth skew line in twistor space, namely the line corresponding to the point at infinity. This introduces many more algebraic invariants which are not actually relevant to the geometry of the transversal itself.

The geometry is much simpler in $\mathbb{CP}^5$, where the four points $x_i$ define a linear subspace. This has an orthogonal subspace (with respect to the inner product defined by $\epsilon_{\alpha\beta\gamma\delta}$), within which just two elements lie on the Klein quadric and so represent points rather than complexified spheres. 

In twistor space, the geometry is particularly elegant. Three lines, say those of $x_1, x_2, x_3$, define a quadric, ruled by two families of lines, each line being transversal to every line of the other family. The line corresponding to  $x_4$ then meets the quadric in just two points. These two points define the two lines which are transversal to all four given lines. The construction is manifestly conformally invariant.
  
Explicitly, the quadric is $\epsilon_{\alpha\beta\gamma\delta}\epsilon_{\lambda\mu\rho\sigma}X_1^{\beta\gamma}X_2^{\delta\lambda}X_3^{\mu\rho}Z^{\alpha}Z^{\sigma}=0$; if $P^{\alpha}$ and $Q^{\alpha}$ are two points on the line $x_4$, then there are just two (possibly coincident) roots $z^{\pm}$ for the twistor $Z^{\alpha} = P^{\alpha} + zQ^{\alpha}$ lying on this quadric; the transversals are then given by $T_{\pm \mu\rho} = \epsilon_{\alpha\beta\gamma\delta}\epsilon_{\lambda\mu\rho\sigma}X_1^{\beta\gamma}X_2^{\delta\lambda}Z_{\pm}^{\alpha}Z_{\pm}^{\sigma}$. By choosing the basis twistors $P^{\alpha}$ and $Q^{\alpha}$ by reference to the line at infinity, one may obtain a formula for the transversal as a point in Minkowski space, but the properties of the transversals do not depend on knowledge of the infinity twistor.

Each transversal is actually a $\mathbb{CP}^1$, and the values $
\kappa, \tilde{\kappa}$ are the  cross-ratios of the points where the four lines intersect it.

The transversal picture illuminates the structure of the degenerate cases where one or more of the external momenta is null.
If $k_i^2=0$ for some $i$, then the quadratic equation $au^2 + (b-a-c)u + c$ has a root at 0, 1, or $ \infty$. Analytically, the dilogarithm in (\ref{eqn:kappa}) is then divergent, and this reflects the fact that the $\mu^2 \rightarrow 0$ limit is no longer finite. In the next section we shall calculate the  form of this limit. But the geometry of the transversals is still well-defined and very simple.

\begin{figure}[h] 
   \centering
   \includegraphics[width=167px]{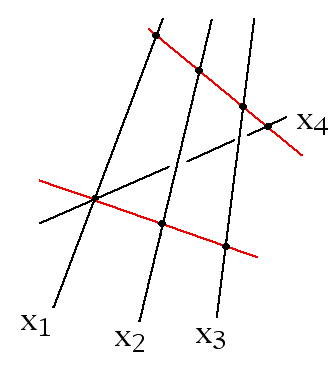} 
   \caption{3-mass transversals}
   \label{fig:Fig.2}
  \end{figure}

 Consider  first the 3-mass case, with $k_1^2=0$, which corresponds to $x_1$ being null-separated from $x_4$, and so to the corresponding lines in twistor space having a common point. The other $K_i$ are non-null, and so all the other lines in twistor space are still skew. There are still just two transversals to all four lines,  but now they are distinguished in the twistor-space picture by a simple geometrical criterion: there is one transversal through the {\em point} common to  $x_4$ and $x_1$, and one in the {\em plane} containing  $x_4$ and $x_1$.

This is quite different from the 4-mass case, where  nothing in the geometry distinguishes one transversal from the other. 
This bifurcation into two types of term, governed by the two possible helicity representations, is very well known. When the box integral is combined with the gauge field structure in order to calculate actual amplitudes, summation over these qualitatively different terms plays an essential part. (Indeed this summation, including the important factor of 1/2, gave rise to the original BCF recursion relation.)  Here we see a natural characterization of this bifurcation in terms of the geometry of transversals in twistor space.

 The momentum-twistor space parameters extend to the degenerate cases very simply. As a matter of convention, we shall allocate names to them in such a way that $K_1$ is associated with $A$ twistors, $K_2$ with $B$ twistors, and so on. In the four-mass case we may take   $x_1, x_2, x_3, x_4$ to correspond to lines $A_2B_1, B_2C_1, C_2D_1, D_2A_1$ respectively. (In applying these results to the actual calculation of $n$-point amplitudes, we would in general have {\em more} than two twistors associated with each $K_i$, but allowing this generalization would not add anything to the analysis at this point.)
 
 For the degenerate cases, the momentum twistors associated with a null momentum simply merge into one. Taking the 3-mass case, we may consider the external momenta as parametrised by {\em seven} momentum twistors $A, B_1, B_2, C_1, C_2, D_1, D_2$, where $x_1, x_2, x_3, x_4$ correspond to lines $AB_1, B_2C_1,$ $ C_2D_1, D_2A$ respectively. Explicitly, we then have
\begin{eqnarray}t = (k_1+K_4)^2 = (x_1 - x_3)^2 = -2 \frac{\langle AB_1C_2D_1\rangle}{\langle AB_1\rangle \, \langle C_2D_1\rangle},\nonumber \\ s = (k_1+K_2)^2 = (x_2 - x_4)^2 = -2 \frac{\langle AB_2C_2D_1\rangle}{\langle B_2C_2\rangle \, \langle D_1A\rangle}\nonumber \\K_2^2 = (x_1 - x_2)^2 = -2 \frac{\langle AB_1B_2C_1\rangle}{\langle AB_1\rangle \, \langle B_2C_1\rangle }, \quad{\rm etc.}  \label{eqn:mompars} \end{eqnarray}
where $ \langle PQRS \rangle$ stands for $\epsilon_{\alpha\beta\gamma\delta}P^{\alpha}Q^{\beta}R^{\gamma}S^{\delta}$ and $ \langle PQ \rangle$ for $I_{\alpha\beta}P^{\alpha}Q^{\beta}$.
  
Although these statements have been made for twistor space, they could equally well have used the {\em dual twistor} representation. In the dual twistor picture, the geometry of the the transversals is interchanged, point becoming plane, and plane becoming point.  

One transversal, that passing through the point $A$, is given by $$T_{\alpha\beta} = \epsilon_{\lambda\mu\pi[\alpha}\epsilon_{\beta] \rho\sigma\tau}A^{\lambda}B_2^{\mu}C_1^{\pi}A^{\rho}D_1^{\sigma}C_2^{\tau}\, .$$ The other, lying in the plane $D_2AB_1$, is given by the dual formula. These formulas also apply to the more degenerate cases. In the `2-mass-easy' case, where $C_1$ and $C_2$ merge into $C$, one transversal is $AC$ and the other its dual.

In the following calculations of the scalar box function, it makes no difference to the technical work, whether the external parameters are momentum twistors or dual momentum twistors. We shall continue by referring to a parametrization with momentum twistors, leaving the dual case implicit.  
 
\section{The box integrals in the degenerate cases}

In the degenerate cases, our problem is that in the $\mu^2\rightarrow 0$ limit, the bi-twistor  $Q$ is simple not just at the vertices of the Feynman tetrahedron, but on from one to four {\em edges}. These singular edges are the source of the logarithmic divergence as $\mu^2\rightarrow 0$.

In principle, one could use this new geometrical setting to re-derive the results of t'Hooft and Veltman (1979) by integrating over a general tetrahedron in the parameter space. This would give the most general picture of the divergence due to the singular edges. But we shall restrict the analysis to the situation of equal $\mu$, and moreover neglect  $O(\mu)$  terms, i.e.\ those which vanish as $\mu^2\rightarrow 0$.  There are many possible representations of the results, but we shall organise them in a form which expresses the $\mu$-dependence entirely in powers of $\log(\mu^2)$, and which facilitates comparison with standard expressions.

We shall use entirely elementary methods (as opposed to Mellin transform methods), with the intention that the location of suitable co-ordinates, and the various devices involving splitting and recombination of terms, may be helpful in further elucidation of the underlying geometric concepts. 

We shall also give an exposition in terms of the standard kinematic variables $s, t, K_i^2$. This is purely for temporary convenience, in that it is easier when working through the details of the integration in the different degenerate cases to be reminded by the notation of the physical setting of null and non-null momenta, and thus  of which quantities have been set to zero before the $\mu^2 \rightarrow 0$ limit is taken. It also assists in making contact with standard expressions. But this notation is in a sense unfortunate and retrograde, because it conceals two important aspects of the theory given here. 

Firstly, each of these variables is defined as a twistor-geometrical invariant of the momentum twistors $Z_i$ and the infinity bi-twistor $I $, by the relations (\ref{eqn:mompars}), and the final value for the integral should likewise be considered as such a twistor-geometrical object, not as a function on standard momentum-space. In particular, the emergence of terms which do not involve $I$, and so are conformally invariant, is a vital feature of the theory. 

Secondly, 
the momentum twistors should all be considered on an equal footing, rather than some being associated with null and some with non-null momenta. In the application to the calculation of loop amplitudes,  a summation is taken over all the ways in which the momentum twistors can clump together into four subsets, and any particular momentum twistor $Z_i$  thus plays many roles within the summation. The complementary paper (Mason and Skinner 2010) uses a notation
which properly illustrates both of these features of the theory, but which is less well suited to the details of integration on which we now embark.

 \subsection{3-mass case}
 
 In what follows, we shall as a matter of convention
take in all cases that $k_1$ is null, i.e. $(x_4-x_1)^2 = 0$, and hence that the corresponding (4-1) edge of the Feynman tetrahedron lies in the singular quadric.

The 3-mass case is the least degenerate, with only that one edge of the tetrahedron on the singular quadric. In this case we can use the same cube-shaped Feynman parameter space defined by (\ref{eqn:cube}), and the integral (\ref{eqn:qintegral}) becomes
\begin{equation}\int_0^{\infty}\int_0^{\infty}\int_0^{\infty}\frac{\dd\alpha\, \dd\beta\, \dd\gamma}{(\alpha(K_2^2+\gamma s) + \beta(t + \gamma K_4^2) + \alpha\beta K_3^2 - \mu^2(1 +\alpha+\beta+\gamma)^2)^2}\, .\label{eqn:3massint}\end{equation}
Note that the $\mu$ only plays a role near $\alpha=\beta=0$, i.e.\ at the edge lying in the singular quadric. Hence we can replace this integral by
  $$\int_0^{\infty}\int_0^{\infty}\int_0^{\infty}\frac{\dd\alpha\, \dd\beta\, \dd\gamma}{(\alpha(K_2^2+\gamma s) + \beta(t + \gamma K_4^2) + \alpha\beta K_3^2 - \mu^2(1 +\gamma)^2)^2}+ O(\mu)$$
\begin{equation}=\int_0^{\infty}\frac{  \log(s+\gamma K_4^2)+ \log(K_2^2 + \gamma t) -  \log(\mu^2K_3^2(1+\gamma)^2)  }{(t+\gamma K_4^2)(K_2^2 + \gamma s)- \mu^2K_3^2(1+\gamma)^2} \dd\gamma+ O(\mu) \, .\label{eqn:3mass2} \end{equation}
The $\mu^2$ term in the denominator can be dropped. This is because the denominator could be expanded as a Taylor series in $\mu^2$, but because, for $n \ge 1$, $(\mu^2)^n \log(\mu^2)$ itself vanishes with $\mu^2$, all terms above zeroth order may be absorbed into the $O(\mu)$. 
The effect is to leave:
 \begin{eqnarray}&&\frac{-\log(-\mu^2K_3^2/K_2^2t)\log(K_4^2 K_2^2/st)}{st-K_2^2K_4^2} \nonumber \\ &+& \int_0^{\infty}\frac{  \log(1+\gamma K_4^2/t)+ \log(1 + \gamma s/K_2^2) - 2 \log(1+\gamma)  }{(t+\gamma K_4^2)(K_2^2 + \gamma s)}\dd\gamma+ O(\mu)\label{eqn:3mass} \, .\end{eqnarray}
  Using the dilogarithmic relation (\ref{eqn:cut}) again, the remaining integral in (\ref{eqn:3mass}) can be written as
  $$\frac{1}{2\pi i}\oint\frac{  \dog(1+\gamma K_4^2/t)+ \dog(1 + \gamma s/K_2^2) - 2 \dog(1+\gamma)  }{(t+\gamma K_4^2)(K_2^2 + \gamma s)}\dd\gamma+ O(\mu)$$  
  $$=\frac{ \dog(1- K_4^2 K_2^2/st)\! -  \!\dog(1- st/K_4^2 K_2^2)\!- \!2\dog(1- K_2^2/s)\! + \!2 \dog(1-t/K_4^2) }{st- K_4^2 K_2^2}$$
Collecting terms, using the dilogarithm identity (\ref{eqn:zz-1}), we obtain:
\begin{eqnarray}\!\!\!\!\!\! &&A^{3m}(s, t, K_2^2, K_4^2, \mu^2) = \nonumber \\ \!\!\!\! \!\!&&\frac{1}{st- K_4^2 K_2^2} \left\{ \begin{array}{l}\log(-\mu^2K_3^2/st)\log(K_2^2K_4^2/st) + 2\dog(1- K_4^2 K_2^2/st) \\ - 2\dog(1-K_2^2/s) - 2\dog(1-K_4^2/t)\\  - {\textstyle \frac{1}{2}}\log^2(K_2^2/s) -{\textstyle \frac{1}{2}} \log^2(K_4^2/t) \end{array}\!\!\! \right\}  .\label{eqn:3massresult}
\end{eqnarray}
This may be written more symmetrically as:
\begin{eqnarray} 
\frac{1}{st- K_4^2 K_2^2}  \left\{ \begin{array}{l}  \log(-\mu^2K_3^2/st)\log(K_2^2K_4^2/st) + 2\dog(1- K_4^2 K_2^2/st)\\ - (\dog(1-K_2^2/s) - \dog(1-t/K_4^2))\\-( \dog(1-K_4^2/t) - \dog(1-s/K_2^2) )
\end{array} \right\}\, . \label{eqn:3massanswer2}
\end{eqnarray}
This expression makes it more transparent that there is no pole when $st= K_2^2 K_4^2$, the analogue of the condition  $\Delta=0$ in the four-mass case.

The full symmetry in $(s,t)\leftrightarrow(K_4^2, K_2^2)$ is manifested by the longer expression:
\begin{eqnarray}\frac{1}{st- K_4^2 K_2^2}  \left\{ \begin{array}{l} \log(-\mu^2K_3^2(stK_2^2K_4^2)^{-\frac{1}{2}})\log(K_2^2K_4^2/st)   \\ + \dog(1- K_4^2 K_2^2/st)- \dog(1- st/K_4^2 K_2^2) \\ -(\dog(1-K_2^2/s) - \dog(1-t/K_4^2))  \\ - ( \dog(1-K_4^2/t) - \dog(1-s/K_2^2) )\end{array}\right\} \, .\label{eqn:3massanswer3}
\end{eqnarray}

In particular, the integral allows $k_1$ not just to be null, but to be {\em zero,} so that $x_4=x_1$. The integral reduces in this case to a triangle defined by the three distinct $x_i$, (The $\gamma$ integral is trivial, as $s=K_2^2, t=K_4^2$.) The calculation above may be considered as treating this configuration as a base and then varying away from it with a non-zero but null $k_1$.

 \subsection{2-mass-easy case}
 
Here $k_3$ is null as well as $k_1$, so two opposite edges of the Feynman tetrahedron lie in the singular quadric. We now have only six  momentum-twistors; i.e.\ $C_1$ and $C_2$ have merged into $C$. There is considerable symmetry: not only is the result  symmetric under $(K_2^2,K_4^2) \leftrightarrow (s,t)$, but also under $K_2^2\leftrightarrow K_4^2$, $s\leftrightarrow t$. There are generalized Feynman parameters which do better at keeping this symmetry manifest, but the following method has the advantage that it requires no solutions of quadratic equations, and also that it extends to the more degenerate cases. It uses the triangular prism (\ref{eqn:prism}), under which the integral becomes
$$ \int_0^{\infty} dx\int \!\int_{\Delta} \frac {\dd\alpha_1 \dd\alpha_2  (1-\alpha_1 -\alpha_2) }
{((1-\alpha_1 -\alpha_2)(\alpha_1(s+K_2^2x) + \alpha_2(K_4^2 + tx)) - \mu^2(1+x))^2
}\, ,$$
where the region $\Delta$ is the triangle $\{0\le \alpha_1, \alpha_2 \le 1, \alpha_1 + \alpha_2 \le 1\}$.
For the next step we give new coordinates to this triangle by  $y = 2(\alpha_1 + \alpha_2), u=4\alpha_1 (1-\alpha_1 - \alpha_2)$. These have the effect of blowing {\em down} the edge  $1-\alpha_1 - \alpha_2=0$ of the triangle into a single point. The result is to map the original tetrahedron into a figure which is a parabolic segment times a line interval, with the original opposite edges becoming two parallel edges at $y=0, y=2$.  The integral becomes
\begin{equation} \int_0^{\infty}\!\!\! \dd x \int_{0}^{2} \!\!\!\dd y \int_{0}^{y(2-y)} \!\!\! \frac {\dd u}
{(u(s+K_2^2x) + ( y(2-y)-u)(K_4^2 + tx) -4\mu^2(1+x))^2}\, .
\label{eqn:2masseasy1}
\end{equation} 
 This is clearly symmetric in $y\leftrightarrow 2-y$, so we
split $ [0 , 2]$ into $[0 , 1] \cup [ 1, 2],$  and change  $y $ to $2-y$ in $[ 1, 2]$. The effect is to concentrate the regularization entirely into the corner at $u=0, y=0$ of the integral
  $$  2\int_0^{\infty}\!\! \dd x \int_{0}^{1} \dd y \int_{0}^{y(2-y)}  \!\!\!\!\frac {\dd u}
{(u(s+K_2^2x) + ( y(2-y)-u)(K_4^2 + tx) -4\mu^2(1+x))^2}\, .$$
Now we use $2 = 2y + (2 - 2y)$ to separate this integral into two pieces:
\begin{equation} 2  \!\!\int_0^{\infty} \!\!\! \dd x \int_{0}^{1}\!\! y \,\dd y \int_{0}^{y(2-y)} \!\! \!\!\!\frac {\dd u}
{(u(s+K_2^2x) +  ( y(2-y)-u)(K_4^2 + tx) -4\mu^2(1+x))^2}\label{eqn:piece1}\end{equation} 
and
\begin{eqnarray}
\! \int_0^{\infty} \!\! \!\dd x \!\int_{0}^{1} \!\!(2-2y)  \dd y \!\!\int_{0}^{y(2-y)} \!\!\!\!\!\!\!\!\!\!\!\!\! \frac {\dd u}
{(u(s+K_2^2x)\! + \!( y(2-y)-u)(K_4^2 + tx) -4\mu^2)^2(1+x))^2}\, .\nonumber\\ \label{eqn:piece2}
\end{eqnarray}
The first piece (\ref{eqn:piece1}) simplifies because it is finite as $\mu \rightarrow 0$, and can be evaluated  immediately as 
$$\frac{2}{st}  \log4 \, \frac{\log(K_2^2 K_4^2/st)}{st -K_2^2 K_4^2 }\, + O(\mu).$$
The second piece (\ref{eqn:piece2}) simplifies because the numerator $(2-2y)$ is just the derivative of $y(2-y)$. Let $w=y(2-y)  - u$ replace $y$, and  it becomes:
$$2 \int_0^{\infty} \dd x\int\!\! \int_{\Delta}  \frac {\dd u \, \dd w}
{(u(s+K_2^2x) + w(K_4^2 + tx) -  4\mu^2(1+x))^2}\, ,$$ 
where $\Delta$ is the triangle $0 \le w, u, w+u \le 1$. 
Elementary  integration of $(u,w)$ over the triangle yields
\begin{eqnarray}
&&2\int_0^{\infty}  \frac{\log(-4\mu^2) +  \log(1+x)}{(s+K_2^2x)(K_4^2 + tx)}\dd x + O(\mu) \label{eqn:piece3}\\
+ &&2\int_0^{\infty} \frac{\log(s+K_2^2x)/(s+K_2^2x) - \log(K_4^2 + tx)/(K_4^2 + tx)}{(s+K_2^2x)-(K_4^2 + tx)} \dd x \, .\label{eqn:piece4}
\end{eqnarray}
The expression (\ref{eqn:piece3}) combines with (\ref{eqn:piece1}) to give   
  $$2 \,\frac{\log(-\mu^2)\log(K_2^2 K_4^2/st) + \dog(1-s/K_2^2) - \dog(1-K_4^2/t)} {st-K_2^2 K_4^2}+ O(\mu).$$
     
 The expression  (\ref{eqn:piece4}) can be rewritten as $$2\int_0^{\infty} \frac{\log s + \log(1+K_2^2x/s)}{(s+K_2^2x)(K_4^2 + tx)} \dd x+ 2\int_0^{\infty} \frac{\log(s+K_2^2x)-\log(K_4^2 + tx)}{(s+K_2^2x)((s+K_2^2x)-(K_4^2 + tx))}\dd x \, .$$
 Of these terms, the first immediately yields
 $$2\, \frac{\log s\log(K_2^2 K_4^2/st) + \dog(1-K_2^2K_4^2/st)}{st-K_2^2 K_4^2 }\,,$$
and for the second term, perform the M\"{o}bius transformation $$w = 1- \frac{s+K_2^2x}{K_4^2 + tx}\, .$$
(This choice turns on the fact that $((s+K_2^2x)-(K_4^2 + tx))^{-1}$ is a removable singularity.  It is tantamount to a proof of Abel's functional equation for the dilogarithm.) It then becomes
\begin{eqnarray}&&\frac{2}{st-K_2^2 K_4^2}  \int_{1-s/K_4^2}^{1-K_2^2/t} \frac {\log(1-w)}{ w (1-w)} \dd w\nonumber \\
 &=& 2\,\frac{ {\textstyle \frac{1}{2}} \log^2(K_2^2/t) -  {\textstyle \frac{1}{2}}\log^2(s/K_4^2) + \dog(1-K_2^2/t) - \dog(1-s/K_4^2) }{st-K_2^2 K_4^2}\, . \end{eqnarray}
 Combining all the terms,  
 we recover the symmetry in $K_2^2$ and $K_4^2$ in the formula:
\begin{eqnarray}&&A^{2me}(s, t, K_2^2, K_4^2, \mu^2) = \nonumber\\
&&\frac{1}{st- K_2^2 K_4^2}  \left\{ \begin{array}{l} 2\log(-\mu^2/\sqrt{st})\log(K_2^2K_4^2/st) + 2\dog(1- K_2^2 K_4^2/st) \\- (\dog(1-K_2^2/s) - \dog(1-t/K_4^2))\\- ( \dog(1-K_4^2/t) - \dog(1-s/K_2^2) )\\
- (\dog(1-K_4^2/s) - \dog(1-t/K_2^2))\\- ( \dog(1-K_2^2/t) - \dog(1-s/K_4^2) )\end{array} \right\}\, .
\end{eqnarray}
The complete symmetry is shown by the longer expression:
\begin{equation}\frac{1}{st- K_4^2 K_2^2}  \left\{\begin{array}{l} \log(\mu^4(stK_2^2 K_4^2)^{-\frac{1}{2}})\log(K_2^2K_4^2/st)   \\ + \dog(1- K_2^2 K_4^2/st) - \dog(1- st/K_4^2 K_2^2) \\ - (\dog(1-K_2^2/s) - \dog(1-t/K_4^2))\\- ( \dog(1-K_4^2/t) - \dog(1-s/K_2^2) )\\- (\dog(1-K_4^2/s) - \dog(1-t/K_2^2))\\- ( \dog(1-K_2^2/t) - \dog(1-s/K_4^2) )\end{array} \right\} \, .\end{equation}
As in the 3-mass integral, this formula shows how the case $K_2^2K_4=st$ is finite. Indeed, it may be seen that the integral allows both $k_1$ and $k_3$ to be zero. In this extreme case  there are only two distinct $x_i$ and only one kinematical invariant, $s=t=K_2^2=K_4^2$. The integral is then $\log(\mu^4/s^2)/s^2$. The method given above may be considered as using this case as baseline and seeing the effect of varying $k_1$ and $k_3$ away from zero.

\subsection{0-mass case}

In the more degenerate cases, where adjacent edges of the Feynman tetrahedron lie on the singular quadric, the integral diverges as $\log^2\mu$ rather than as $\log \mu$, and more intricate work is necessary to separate the singular part.

The most degenerate case is that of the 0-mass integral. Then there are just four momentum twistors $A, B, C, D.$ The parameters
$x_1, x_2, x_3, x_4$ correspond to lines $AB, BC, CD,$ $ DA$ respectively. 

We may take from the preceding analysis the expression (\ref{eqn:2masseasy1}), and set $K_2^2 = K_4^4 = 0$. In this case there is no point in deferring the $x$-integration, after which it becomes 
$$   \int_{0}^{1} 2\, \dd y \int_{0}^{y(2-y)}  \frac {\dd u}
{(us - 4\mu^2)( (y(2-y) - u)t  - 4\mu^2)} \, ,$$
which again we split into two pieces again by writing $2=2y +( 2-2y)$.

The first piece is not now infra-red finite, but can still be easily evaluated as:
\begin{eqnarray}
&&=   \frac{2}{st}  \int_{0}^{1}  \frac{ 2\log( y(2-y)) + \log(-4\mu^2/s) + \log(-4\mu^2/t)}{2-y} \,\dd y  \,  + O(\mu) \nonumber \\
&&= \frac{2}{st} \{(\log 2)^2 - \pi^2/12  - ( \log(-4\mu^2/s) + \log(-4\mu^2/t))\log 2 + O(\mu) \, .\end{eqnarray}

The second piece is: 
\begin{eqnarray}&&\frac{2}{st} \int \!\! \int_{\Delta}   \frac {\dd u \, \dd w}
{(u - 4\mu^2/s)(w  - 4\mu^2/t)}\nonumber \\
&=&  \frac{2}{st}\{\log(- 4\mu^2/s)\log(- 4\mu^2/t) - \pi^2/6\} + O(\mu)\, . \end{eqnarray}
Combining the terms, we have 
\begin{equation}A^{0m}(s, t, \mu^2)= \frac{1}{st} \{ 2 \log( -\mu^2/s) \log (-\mu^2/t) - \pi^2 \} +O(\mu)\, . \label{eqn:0massanswer}\end{equation}

\subsection{1-mass case}

When $K_4^2 = 0,$ but $K_2^2 \ne 0$, the first piece of the integral becomes
$$2\,  \int_{0}^{1} y \, \dd y \int_{0}^{y(2-y)}  \frac {\dd u}
{(y(2-y)K_2^2 + (t-K_2^2)u -4\mu^2)( (y(2-y) - u)s  -4\mu^2)}$$
Integrate out $u$, and obtain
\begin{eqnarray}&&\frac{2}{st}\int_{0}^{1} \dd y \frac{2\log (y(2-y)) - \log(-4\mu^2K_2^2/st)}{2-y} + O(\mu)\nonumber \\
&=& \frac{2}{st}(\log 4 \log(-4\mu^2K_2^2/st) +(\log2)^2 - \pi^2/12) + O(\mu)\, .\end{eqnarray}
The second piece is
$$2  \int\!\! \int_{\Delta}  \frac {\dd u \, \dd w}
{(us + wK_2^2 -  4\mu^2)( wt -  4\mu^2)}\, ,$$ 
where $\Delta$ is the triangle $0 \le w, u, w+u \le 1$. Integrating out $u$, this is
$$\frac{2}{s}  \int_{0}^{1} \frac{\dd w}{wt-4\mu^2} \log(1-w(1-K_2^2/s) - 4\mu^2/s)-\log(wK_2^2/s - 4\mu^2/s) \, .$$
The first logarithm is infra-red finite and integrates immediately to $$ \frac{2}{st}  \dog(1-K_2^2/s) + O(\mu)\, . $$
In the second logarithm
substitute $x=-4\mu^2/(wt -4\mu^2)$ and hence obtain
\begin{eqnarray}&& \frac{2}{st} \int_{-4\mu^2/t}^{1} \frac{\dd x}{x}\{ \log( -4\mu^2 K_2^2/stx )  + \log(1 - x(1-t/K_2^2)) \}+ O(\mu) \nonumber \\
&=& \frac{2}{st}  \left( \begin{array}{l}( \log(-4\mu^2/t)\log( -4\mu^2 K_2^2/st) - {\textstyle \frac{1}{2}}\log^2(-4\mu^2/t) \\+ \dog(1-t/K_2^2) ) + O(\mu) \end{array} \right) \, . \end{eqnarray}
These terms combine to give:
\begin{eqnarray}A^{1m}(s, t, K_2^2, \mu^2)= \frac{1}{st} \left\{ \begin{array}{l} 2\log( -\mu^2/s) \log (-\mu^2/t) -  \log^2(-\mu^2/K_2^2) \\ - 2\dog(1-K_2^2/s)  - 2\dog(1-K_2^2/t)  - \pi^2/3  \\ +O(\mu) \end{array} \right\} . \label{eqn:1massanswer}
\end{eqnarray}

\subsection{2-mass-hard case}

For the `2-mass-hard' case, with $K_4$ and $K_3$ non-null, we have six momentum twistors 
$A, B, C_1, C_2, D_1, D_2$. A little more ingenuity is required for this integral. The method given here is probably not optimal, but gives an indication of the geometrical relationships underlying the structure of the result.  We start with the fully symmetrical Feynman tetrahedron as the parameter space, so that the integral is:
$$ \int_0^1\int_0^1\int_0^1\int_0^1\frac { \delta(\alpha_1 + \alpha_2 + \alpha_3 +\alpha_4 - 1) \,\, \dd\alpha_1 \dd\alpha_2 \dd\alpha_3 \dd\alpha_4}
{(\alpha_1\alpha_3 t + \alpha_2\alpha_4 s  + \alpha_2\alpha_3 K_3^2 + \alpha_3\alpha_4 K_4^2 - \mu^2)^2
}\, .$$
Note that symmetry between $x_2$ and $ x_3$ implies a symmetry between $K_3^2$ and $K_4^2$. Eliminating $\alpha_4$ and $\alpha_1$, this is
\begin{eqnarray} \int\!\!\!\int_{\Delta} \frac {(1-\alpha_2 - \alpha_3)\,\, \dd\alpha_2 \dd\alpha_3  }
{\left( \begin{array}{l} (\alpha_2(1-\alpha_2-\alpha_3)s + \alpha_2 \alpha_3 K_3^2 + \alpha_3(1-\alpha_2-\alpha_3)K_4^2-\mu^2)\\
( \alpha_3(1-\alpha_2-\alpha_3)t  +\alpha_2 \alpha_3 K_3^2  -\mu^2)\end{array}\right)} \,,
\end{eqnarray}
where the region of integration $\Delta$ is the triangle $0<\alpha_2, \alpha_3 < 1, \alpha_2 + \alpha_3 < 1$.

 Now this integral, but with $K_4^2 = 0$, has already been evaluated, as it is a 1-mass integral.
We therefore study the {\em difference} between the 2-mass-hard integral and this 1-mass integral.  
This difference  is
\begin{eqnarray} \int\!\!\!\int_{\Delta} \frac { \alpha_3\, K_4^2\, (1-\alpha_2 - \alpha_3)^2 \,\, \dd\alpha_2 \dd\alpha_3  }
{\left( \begin{array}{l}(\alpha_2(1-\alpha_2-\alpha_3)s + \alpha_2 \alpha_3 K_3^2-\mu^2) \\
( \alpha_2(1-\alpha_2-\alpha_3)s + \alpha_3(1-\alpha_2-\alpha_3)K_4^2 +\alpha_2 \alpha_3 K_3^2 -\mu^2)
\\( \alpha_3(1-\alpha_2-\alpha_3)t +\alpha_2 \alpha_3 K_3^2 -\mu^2)\end{array} \right) }\, .
\end{eqnarray}
Next, we use a splitting of the numerator factor $$\alpha_3(1-\alpha_2-\alpha_3) = [- \alpha_2 \alpha_3 K_3^2/t] +   [\mu^2/t ]+ [(\alpha_3(1-\alpha_2-\alpha_3)t ++ \alpha_2\alpha_3K_3^2 - \mu^2)/t] $$ to divide this integral into three pieces. The first piece is infra-red finite, for up to $O(\mu)$ terms it is:
\begin{eqnarray} -\frac{K_3^2 K_4^2}{t} \int\!\!\!\int_{\Delta} \frac {(1-\alpha_2 - \alpha_3) \,\,\dd\alpha_2 \dd\alpha_3 }
{\left( \begin{array}{l}((1-\alpha_2-\alpha_3)s +  \alpha_3 K_3^2)\\
( \alpha_2(1-\alpha_2-\alpha_3)s + \alpha_3(1-\alpha_2-\alpha_3)K_4^2 +\alpha_2 \alpha_3 K_3^2 ) \\
( (1-\alpha_2-\alpha_3)t +\alpha_2  K_3^2 )\end{array}\right)} \, .
\end{eqnarray}
To evaluate this, it is convenient to change co-ordinates to:
$$y = (1-\alpha_2 - \alpha_3)/(\alpha_2 K_3^2), \quad v = (1-\alpha_2 - \alpha_3)/(\alpha_3 K_3^2),$$yielding
\begin{eqnarray}&&\frac{K_4^2}{t}\int_0^{\infty} \!\!\int_0^{\infty} (1+zt)^{-1}(ys + K_4^2 z + 1)^{-1}(1+ys)^{-1} \dd y \, \dd z \label{eqn:irpiece} \\
= &&\frac{1}{st}\int_0^{\infty} z^{-1}(1+zt)^{-1}\log(1 + K_4^2z)\, \dd z = -\frac{1}{st} ( \dog(1- K_4^2/t) - \pi^2/6)\nonumber \end{eqnarray}
by using equation {(\ref{eqn:cut1}) in the Appendix.

The second piece is
\begin{eqnarray}&& \int\!\!\!\int_{\Delta} \frac { \mu^2 K_4^2\, (1-\alpha_2 - \alpha_3) \,\, \dd\alpha_2 \dd\alpha_3  }
{\left( \begin{array}{l}(\alpha_2(1-\alpha_2-\alpha_3)s + \alpha_2 \alpha_3 K_3^2-\mu^2) \\
( \alpha_2(1-\alpha_2-\alpha_3)s + \alpha_3(1-\alpha_2-\alpha_3)K_4^2 +\alpha_2 \alpha_3 K_3^2 -\mu^2)
\\( \alpha_3(1-\alpha_2-\alpha_3)t +\alpha_2 \alpha_3 K_3^2 -\mu^2)\end{array} \right) }\\
 =&&K_4^2\, \int\!\!\!\int_{\Delta} \frac { \mu^2 \,  \dd\alpha_2 \dd\alpha_3  }
{(\alpha_2s -\mu^2) 
( \alpha_2s + \alpha_3K_4^2 -\mu^2)
( \alpha_3 t  -\mu^2) }\,  + O(\mu),
\end{eqnarray}
which is essentially the same integral as (\ref{eqn:irpiece}), with $\mu^2$ playing the role of $K_3^2$, and again gives
$$ -\frac{1}{st} ( \dog(1- K_4^2/t) - \pi^2/6)\, .$$
The third piece is
\begin{equation} \frac{K_4^2 }{t}\int\!\!\!\int_{\Delta} \frac {(1-\alpha_2 - \alpha_3) \,\, \dd\alpha_2 \dd\alpha_3}
{ \left( \begin{array}{l} 
( \alpha_2(1-\alpha_2-\alpha_3)s + \alpha_3(1-\alpha_2-\alpha_3)K_4^2 +\alpha_2 \alpha_3 K_3^2  - \mu^2 )
 \\
( \alpha_2(1-\alpha_2-\alpha_3)s +\alpha_2 \alpha_3 K_3^2 -\mu^2)
\end{array}\right)} \, , \end{equation}
which we may write as
\begin{eqnarray}
&&\frac{K_4^2 }{t}\int\!\!\!\int_{\Delta} \dd\alpha_2 \dd\alpha_3 \int_0^{1-\alpha_2-\alpha_3}\!\!\!\frac{ \dd\alpha_1}{(\alpha_2(1-\alpha_2-\alpha_3)s + \alpha_1\alpha_3K_4^2 +\alpha_2 \alpha_3 K_3^2  - \mu^2 )^{2}} \nonumber \\
&=&\frac{K_4^2 }{t}\int_0^{1}\!\!\int_0^{1}\!\!\int_0^{1} \!\!\int_0^{1}\!\frac{\delta(\alpha_1+\alpha_2+\alpha_3+\alpha_4-1)\,  \dd\alpha_1\,\dd\alpha_2\, \dd\alpha_3\, \dd\alpha_4 }{(\alpha_2(\alpha_1 + \alpha_4)s + \alpha_1\alpha_3K_4^2 +\alpha_2 \alpha_3 K_3^2  - \mu^2 )^{2}}\nonumber \\
&=&\frac{K_4^2}{t}A^{2mh}(K_4^2, s, s, K_3^2, \mu^2) \quad =\quad  \frac{K_4^2}{t}A^{2mh}(K_4^2, s, K_3^2, s, \mu^2)\end{eqnarray}
This object can be considered as the 2-mass-hard amplitude obtained from the one we are considering by merging $k_1$ and $k_2$ into a single non-null momentum $k_1 + k_2$, but splitting $K_4$ into a sum of two null momenta, one of them orthogonal to $k_1 + k_2$. 

In what follows it is convenient to use the dimensionless functions
\begin{equation}f_2(s, t, K_2^2, K_4^2) = stA^{2mh}(s, t, K_2^2, K_4^2), \quad f_1(s, t, K_2^2) = stA^{1m}(s,t,K_2^2)\, ,\label{eqn:fa}\end{equation} which are just twice the conventional `box functions'.  In these terms
(suppressing the dependence on $\mu$ and the $O(\mu)$ terms temporarily)
 we have the identity:
\begin{eqnarray}f_2(s, t, K_3^2, K_4^2) = f_1(s, t, K_3^2) - f_2(K_4^2, s, s, K_3^2) -2(\dog(1-K_4^2/t) - \pi^2/6) \label{egn:identity} \end{eqnarray}
By applying this identity three times we can find $f_2$ without any more integration. We have  \begin{eqnarray}f_2(K_4^2, s, s, K_3^2)\! \!&=&\! \!f_1(K_4^2, s, s)  - f_2(K_3^2, K_4^2, K_4^2, s)-2(\dog(1-K_3^2/s) \!- \!\pi^2/6)\, ,\nonumber \\
f_2(K_3^2, K_4^2, K_4^2, s)\! \! &=&\! \!f_1(K_3^2, K_4^2, K_4^2) \! -\! f_2(s, K_3^2, K_3^2, K_4^2)\!-\!2 (\dog(1\!-\!s/K_4^2)\! - \!\pi^2/6)\, , \nonumber\\
f_2(s, K_3^2, K_3^2, K_4^2)\!\!  &=&\! \!f_1(s, K_3^2, K_3^2) - f_2(K_4^2, s, s, K_3^2)  -2( \dog(1\!-\!K_4^2/K_3^2) \!-\! \pi^2/6)\, .\nonumber 
\end{eqnarray}
Hence
\begin{eqnarray}2 f_2(K_4^2, s, s, K_3^2) &=& f_1(K_4^2, s, s) -2 (\dog(1-K_3^2/s) - \pi^2/6)\, .\nonumber \\
&-&f_1(K_3^2, K_4^2, K_4^2) +2 (\dog(1-s/K_4^2) - \pi^2/6) \nonumber\\
&+&f_1(s, K_3^2, K_3^2) -2 (\dog(1-K_4^2/K_3^2)- \pi^2/6)\, .
\end{eqnarray}
Using the known form of the $f_1$ function, from (\ref{eqn:1massanswer}), this yields 
\begin{eqnarray}2f_2(K_4^2, s, s, K_3^2)
&=& -4 \dog(1-K_3^2/s) +  \log^2(-\mu^2/s) + \log^2(-\mu^2/K_4^2) \nonumber \\& &-\log^2(-\mu^2/K_3^2) - 2\log(K_3^2/s)\log(K_4^2/s) \, .\end{eqnarray}
Hence
\begin{eqnarray} A^{2mh}(s, t,  K_3^2, K_4^2, \mu^2) =\frac{1}{st}\left\{\begin{array}{l} 2 \log(-\mu^2/s)\log(-\mu^2/t) -\textstyle{\frac{1}{2}} \log^2(-\mu^2/s)\\-\textstyle{\frac{1}{2}} \log^2(-\mu^2/K_3^2)-\textstyle{\frac{1}{2}} \log^2(-\mu^2/K_4^2) \\ +  \log(K_3^2/s)\log(K_4^2/s)\\   - 2\dog(1-K_3^2/t) -  2\dog(1-K_4^2/t)\\
+ O(\mu)\end{array}  \right\} \, .
\end{eqnarray}
Note that the symmetry in $K_3^2$ and $K_4^2$ is restored.
It would be advantageous to see how these transformations and identities relate to  geometric regions which break up the tetrahedron, and deal with the singular edge and vertex effects.

\section{Generalization to independent $\mu_i$}

The 3-mass integral is the one degenerate case where it is simple to generalize from a single $\mu^2$ to independent $\mu_1^2, \mu_2^2, \mu_3^2, \mu_4^2$ in the obvious sense. We follow
 the same analysis as in section 7.1, and the only difference from (\ref{eqn:3massint}) is that the term $\mu^2(1+\alpha+\beta+\gamma)^2$ is replaced by $(\mu_1^2 + \alpha\mu_2^2 + \beta\mu_3^2 +\gamma\mu_4^2)(1+\alpha+\beta+\gamma)$.
 
By the same arguments the resulting integral may be simplified, up to terms which vanish as $\mu_i \rightarrow 0$, to
$$\int_0^{\infty}\!\int_0^{\infty}\!\int_0^{\infty}\!\frac{\dd\alpha\dd\beta \dd\gamma}{(\alpha(K_2^2+\gamma s) + \beta(t + \gamma K_4^2) + \alpha\beta K_3^2 - (1 +\gamma)(\mu_1^2  +\gamma\mu_4^2))^2}\, .$$
Notably, $\mu_2$ and $\mu_3$ can be taken to zero; finiteness only requires regulators on those propagators which meet the null momentum.

The integral will thus differ from (\ref{eqn:3mass2}) only in that $\log(\mu^2(1+\gamma)^2)$ is replaced by $\log((\mu_1^2 + \gamma \mu_4^2)(1 + \gamma))$, and so the result of the integral differs from the stated value of the 3-mass integral (\ref{eqn:3massresult}) by
$$ \int_0^{\infty}\frac{  \log((\mu_1^2+\gamma \mu_4^2)/(\mu^2(1 + \gamma) )  }{(t+\gamma K_4^2)(K_2^2 + \gamma s)}\, \dd\gamma \, .$$
This is a perfectly well-defined function of the kinematic scalars and $\mu_1, \mu_2$, but the generalization of our asymptotic formulas is more problematic, because statements about limiting functions depend strongly on exactly how the limits  $\mu_1, \mu_4 \rightarrow 0$ are taken.  
 If, for instance, the ratio $\mu_4/\mu_1$ tends to some number $\lambda$ then this difference effect is
\begin{eqnarray} \frac{1}{st- K_4^2 K_2^2}\left\{ \begin{array}{l} \{ \log(\mu_1\mu_4/\mu^2)\log(K_2^2K_4^2/st)\nonumber \\ - \dog(1-\lambda K_2^2/s) + \dog(1-K_2^2/s)\\ + \dog(1-\lambda t/K_4^2) - \dog(1-t/K_4^2) \end{array}  \right\}\, .
\end{eqnarray}
We might like to make a comparison with the formalism of  Alday,  Henn, Plefka and Schuster (2009), which treats the effect of different mass parameters. But it is important to note that for different mass parameters their limit takes a quite different form from that discussed above. In their formalism, the external null momentum $k_1$ is approached as a limit in which $K_1^2 = (\mu_1 - \mu_4)^2$. Such a limit is expressible in the formalism described in this note, but it is not very natural in a twistor-geometric setting, as there is no continuous transition from a small  mass to zero mass.  Either one momentum twistor is needed, or more than one.

\section{Summary}

The results given above are in agreement with standard results obtained by dimensional regularization methods, as tabulated by Bern et al.\  (1993), Bern et al.\ (2004), in the following sense. First, allowance must be made for the fact that we have given the box {\em integrals} rather than the box {\em functions}, as described in section 2 above. The correspondence may then be obtained by multiplying the dimensionally regularized expression by $(\mu^2)^{\epsilon}$, subtracting simple and double poles in $\epsilon$, and then taking $\epsilon = 0$.

We have neglected terms which vanish with $\mu^2$, but these have an important singularity structure; they cannot be thought of as referring to a Taylor series in $\mu^2$. Their  finiteness has the character of a finite value at a branch point, like $z\log z, \log(1-z)\log z$ or $  \dog(1-z)$   at $z=0$. It would also be more correct to write them as  $O(\mu^2/M)$, where $M$ is the minimum of the kinematical variables $s, t, K_i^2$, to indicate that the approximation of these divergent functions by functions involving  $\log \mu^2$ is no approximation at all unless $\mu^2$ is small compared with all the other parameters. This  gives another way of seeing why the various limits $\mu^2 \rightarrow 0, K_i^2 \rightarrow 0$ do not commute, so that each degenerate case has to be considered separately.

The various expressions for the box integrals show three levels of invariant structure: there are divergent terms which are not even scale-invariant, there are  terms like $\dog(1-K_2^2/s)$ which  are scale-invariant but break conformal invariance, and there are  conformally invariant terms like  $\dog(1-K_2^2K_4^2/st)$. When the expressions are written in the natural momentum-twistor parameters, these features are immediately apparent.  In the case of a {\em six}-field amplitude, for instance, when the expressions above are written in terms of six momentum-twistors $Z_1\ldots Z_6$, the term $\dog(1-K_2^2K_4^2/st)$ becomes $\dog(1- \langle 1234\rangle \langle 4561\rangle/  \langle 6134 \rangle \langle 4512\rangle )$ = $\dog( \langle 1345\rangle \langle 1246 \rangle/  \langle 6134 \rangle \langle 4512\rangle )$, where $\langle ijkl\rangle = \epsilon_{\alpha\beta\gamma\delta} Z_i^{\alpha} Z_j^{\beta}Z_k^{\gamma}Z_l^{\delta}$, thus showing the manifest conformal symmetry. In contrast, the breaking of conformal invariance in the other terms  is shown explicitly by their dependence on the infinity bi-twistor $I$. At this point we refer to the complementary work of Mason and Skinner (2010) which considerably develops this geometrical structure. Generally, it would appear that this exploration of the twistor geometry of the box functions gives only a first hint of new structures which are emerging from the analysis.

\section{Historical note} Much interest in early twistor theory lay in the study of the integral
\begin{equation}\int \frac{1}{(x-p_1)^2 (x-p_2)^2(x-p_3)^2 (x-p_4)^2 } \,\, \dd^4x\label{eqn:phi4}\end{equation}
arising in first-order
 $\phi^4$ theory, i.e.\ in $x$-space, not in momentum space. Here the $x$-integral is over real $x$, and the $p_i$ are four points in $\mathbb{CM}$, of which two must be in the future tube and two in the past tube for a non-singular, non-vanishing result. Thus, $1/(x-p_1)^2$ is a classical positive or negative frequency solution of the massless scalar field, called an `elementary state' by twistor theorists. The choices between future and past tube gives rise to three `channels', which correspond to the three logarithmic period functions noted in (\ref{eqn:logperiod}).  (For instance, the first of those logarithmic functions allows $a=c, b=0$, which corresponds to allowing $p_1=p_2, p_3=p_4$.) The first twistor-integration results were stated by Penrose and McCallum (1972); curiously, these mis-identified the analogue of the `leading singularity' contour as the correct (`logarithmic') contour for an amplitude. Thus even in the earliest days, these questions of connecting contour with physical amplitude played a dominating role in twistor integration. This was soon corrected, as later surveys, such as by Hodges and Huggett (1980), Hodges (1983), indicate. 

The construction of the dilogarithmic result for the 4-mass integral by an $S^7 \times S^1$ contour appeared in an informally published note (Hodges 1977).  Its motivation was different from that presented in this note; its title was `Crossing and twistor diagrams' and its hope was to define a `super-amplitude' for the $\phi^4$ integral (\ref{eqn:phi4}), in such a way that each channel would arise as a period. The construction used the linearity of $\mathbb{CP}^5$, leading to the basic dilogarithmic answer, and its relationship to the logarithmic periods. The statement of the result did not fully address the problem that the dilogarithmic result can only be regarded as a limiting value as elements of $\mathbb{CP}^5$ approach the singular quadric. The original `crossing' idea could not be generalized from this application to elementary states, and was never taken any further.  This present note arose in the context of the renaissance of twistor field theory in the wake of Witten's twistor string model in 2003.  It then became notable that (a) the dilogarithmic form of the 4-mass box integral is just that of the structure found in 1977 and (b) that the most obvious way to make the 1977 construction rigorous is to add the $\mu^2 I^{\alpha\beta}$ terms, and regard the finite result as a limit as $\mu^2\rightarrow 0$. 

Use of a generalized `Feynman trick' was widespread in early twistor theory work, following Penrose's use of it in many examples. As an example, the study of the $\phi^4$ scalar integral (\ref{eqn:phi4}) in (Hodges 1983) used the fact that the product of fields $1/(x-p_1)^2(x-p_2)^2$ can be written as a Dirichlet average of fields of the simpler form $1/((x-z)^2)^2$, where $z$ varies over a {\em sphere} in $\mathbb{CM}$.

The 1977 note also used the fact that the contour for the fundamental twistor integral (\ref{eqn:fundamental}) need not be realised as an $S^7\times S^1$. Instead, it integrated $W_{\alpha}$ as a $S^1$ overall phase, times a tetrahedron in projective $W_{\alpha}$ space, with vertices at $W_{\alpha}=A_{\alpha}, B_{\alpha}, C_{\alpha}, D_{\alpha}$. The result is $$(2\pi i)^{-4}\oint \epsilon^{\alpha\beta\gamma\delta}A_{\alpha}B_{\beta}C_{\gamma}D_{\delta} ((Z^{\alpha}A_{\alpha})(Z^{\alpha}B_{\beta})(Z^{\gamma}C_{\gamma})(Z^{\delta}D_{\delta} ))^{-1}\dd^4Z \,.$$ A direct product of four $S^1$ integrals in $Z$ completes the integration. This contour is actually homologous to the $S^7\times S^1$ when four extra regions are added, each of which is inside a space of form $W_{\alpha} = \mathrm{constant}$, and so contributes nothing to the integral.
The use of this contour makes the basic Feynman loop integral (\ref{eqn:basic}) trivial, if the tetrahedron is chosen with vertices on $O$ and $I$. 

The motivation for introducing this complication was that whilst the $S^7 \times S^1$ contour clearly did not exist in the singular limit, some adaptation of this boundary-defined contour might in fact survive in the limiting case. This idea may not be completely wrong, in view of the central role of boundary-defined integrals in momentum-twistor space (Hodges 2009). In any case, it is notable that contour integration problems of the 1970s are very relevant today.

\section{Acknowledgments}
This investigation was considerably stimulated by a workshop meeting at the Mathematical Institute,  University of Oxford, in June 2009. The distinguished participants included both Roger Penrose and Nima Arkani-Hamed.  At this meeting, James Drummond pointed out the subtle breaking of dual conformal symmetry which occurs even in the 4-mass case, and supplied the reference to the work of Duplan\u{c}i\'{c} and Ni\u{z}i\'{c} (2002). Lionel Mason made a particular valuable contribution on the fibration of $S^7$. He and David Skinner, in developing the connection of this material with $\mathbb{P}^5$ geometry, have supplied  ever-helpful comment.   
I also had the advantage of being able to discuss some of these ideas at a Workshop on Hidden Structures in Field Theory Amplitudes at the Niels Bohr Institute, Copenhagen, 12-14 August 2009, and of hearing from Johannes Henn about his group's parallel application of five-dimensional geometry and mass-parameter regularization.  This note was completed during a visit to the Institute for Advanced Study, Princeton, thanks to an invitation from Nima Arkani-Hamed, and the generous support provided by that institution. It has benefited from discussion in this highly stimulating environment.

Long ago, it was funding from the the former Science and Engineering Research Council which gave me the opportunity to explore the twistor geometry of scattering integrals, and it is a pleasure to observe that, albeit somewhat late in the day, the material in which the SERC invested now attracts greater interest than it did in the 1970s. But the specific suggestion to study dilogarithms came directly from Roger Penrose in 1977, and his  insight into conformal geometry  underlies everything in this note.  

\section{Appendix}

The dilogarithm function $\dog z$, also written $\mathrm{Li}_2(z)$, is defined by
\begin{equation}\dog\, z =- \int_{0}^{z}\frac{\log(1-w)}{w} \dd w = \sum_1^{\infty} \frac{z^n}{n^2} 
\, .\label{eqn:dilog}
\end{equation}
By elementary integration,
\begin{eqnarray}\dog(1-z^{-1}) &=& -\dog(1-z) -\frac{1}{2}\log^2 z \, ,\\
\dog(1-z)& =& -\dog \,z + \frac{\pi^2}{6} - \log(1-z)\log z \label{eqn:zz-1}
\\
\dog(z^{-1}) &=& -\dog \,z - \frac{\pi^2}{6} -\frac{1}{2} \log^2(-z)\, ,\end{eqnarray}
showing how $\dog \,z$ behaves under cross-ratio transformations.  At $z=1$, $\dog\, z$ is not analytic, and has a branch point. But it also has the finite value $\pi^2/6$  there. It is useful to consider $\dog \,z$ as defined on the complex plane with a cut from 1 to $\infty$ along the real axis; it then has a discontinuity of $2\pi i \log z$ across this cut.
It follows that the definite integral
$$  \int _0^{\infty}
\frac{\log(1+\lambda z)}{(z+\kappa)(z+ \tilde{\kappa})} \dd z $$
is equivalent to the contour integral 
\begin{equation} \frac{1}{2\pi i}  \oint 
\frac{\dog(1+\lambda z)}{(z+\kappa)(z+ \tilde{\kappa})} \dd z\,= \frac{\dog(1-\lambda\kappa) - \dog(1-\lambda\tilde{\kappa})}{\tilde{\kappa}- \kappa} , \label{eqn:cut} \end{equation}
and we use this repeatedly. Here $\kappa$ and $\tilde{\kappa}$ are assumed to be away from the cut.  But by taking the limit  $\tilde{\kappa}\rightarrow 0$, we deduce
\begin{equation}  \int _0^{\infty}
\frac{\log(1+\lambda z)}{(z+\kappa)z} \, \dd z = - \frac{\dog(1-\lambda\kappa) - \pi^2/6}{\kappa} \,.
\label{eqn:cut1}
\end{equation}

\section{References}

L.\ F.\ Alday,  J.\ Henn, J.\ Plefka and T.\ Schuster, Scattering into the fifth dimension of $N=4$ Yang-Mills, arXiv 0908.0684 (2009)

N.\ Arkani-Hamed, F.\ Cachazo and C.\ Cheung, The Grassmannian origin of dual superconformal invariance, 
arXiv 0909.0483v1 (2009)

Z.\ Bern, L.\ Dixon and D.\ A.\ Kosower, Dimensionally regulated pentagon integrals, arXiv:hep-ph/9306240 (1993)

Z. Bern, V. Del Duca, L. J. Dixon and D. A. Kosower, All non-maximally-helicity-violating one-loop seven-gluon amplitudes in $N = 4$ super-Yang-Mills theory, arXiv:hep-th/0410224 (2004)

 G.\ Duplan\u{c}i\'{c} and B.\ Ni\u{z}i\'{c}, IR finite one-loop box scalar integral with massless internal lines, arXiv:hep-ph/0201306 (2002) 

A.\ Hodges, Crossing and twistor diagrams, in {\em Twistor Newsletter} 5, 11 July 1977, Mathematical Institute, University of Oxford, reprinted in {\em Further Advances in Twistor Theory}, eds.\ L.\ P.\ Hughston and R.\ S.\ Ward (Pitman, 1979)

A.\ Hodges and S.\ A.\ Huggett, Twistor diagrams, {\em Surveys in High Energy Physics} {\bf 1}, 333-353 (1980)

A. Hodges, Twistor diagrams and massless M\"{o}ller scattering, {\em Proc. R. Soc. Lond.} {\bf A 385,} 207-228 (1983)

A. Hodges, Eliminating spurious poles from gauge-theoretic amplitudes, arXiv 0905.1473v1 (2009)

L.\ Mason and D.\ Skinner, Dual superconformal invariance, momentum twistors and Grassmannians, arXiv 0909.0250v1 (2009)

L.\ Mason and D.\ Skinner, Amplitudes at weak coupling as polytopes in $\mathrm{AdS}_5$, to appear (2010)

R.\ Penrose and M.\ A.\ H.\ McCallum, Twistor theory: an approach to the quantisation of fields and space-time, {\em Physics Reports} {\bf 4}, 241 (1972)

\end{document}